\documentclass[prb,preprint]{revtex4-1}
\usepackage{amsmath}  % needed for \tfrac, \bmatrix, etc.
\usepackage{amsfonts} % needed for bold Greek, Fraktur, and blackboard bold
\usepackage{graphicx} % needed for figures
\newcommand{\beq}{\begin{equation}}
\newcommand{\eeq}{\end{equation}}
\newcommand{\beqn}{\begin{eqnarray}}
\newcommand{\eeqn}{\end{eqnarray}}

\newcommand{\hti}{{\widetilde{H}}}
\newcommand{\kti}{{\widetilde{K}}}
\newcommand{\brarpa}{{\langle 0 |}}
\newcommand{\ketrpa}{{| 0 \rangle}}
\newcommand{\brao}{{\langle 1 |}}
\newcommand{\keto}{{| 1 \rangle}}
\newcommand{\brahf}{{\langle {\rm HF} |}}
\newcommand{\kethf}{{| {\rm HF} \rangle}}
\newcommand{\erpa}{E^{\rm RPA} }
\usepackage[usenames]{color}
\usepackage{ulem}

\begin{document}

\title{Hartree-Fock and Random Phase Approximation theories\\
in a many-fermion solvable model}

\author{Giampaolo Co',}\,
\email{gpco@le.infn.it}
\affiliation{Dipartimento di Matematica e Fisica ``Ennio De Giorgi",\\
Universit\`a del Salento, Lecce,\\ and \\
INFN, sez.\,di Lecce, Italia,}
% Please provide a full mailing address here.

\author{Stefano De Leo,}
\email{deleo@ime.unicamp.br}
\affiliation{Departamento de Matem\'atica Aplicada,\\ Universidade Estadual de Campinas,\\ Campinas, Brasil.\\
{\sl [\,Modern Physics Letter A {\bf 30}, 1550196-15 (2015)\,]}}

% See the REVTeX documentation for more examples of author and affiliation lists.

%\date{Prova}

\begin{abstract}
We present an ideal system of interacting fermions where the solutions
of the many-body Schr\"odinger equation can be obtained without making
approximations. These exact solutions are used to test the validity of two
many-body effective approaches, the Hartree-Fock and the Random Phase
Approximation theories. The description of the ground state done by the
effective theories improves with increasing number of particles.
\end{abstract}
% AJP requires an abstract for all regular article submissions.
% Abstracts are optional for submissions to the "Notes and Discussions" section.

\maketitle % title page is now complete

\section{Introduction}
The difficulty of solving the Schr\"odinger equation for quantum
many-body systems has
induced the development of theories based on approximations which simplify the
problem \cite{mes61,fet71,nol09}.
The formalism of these theories
is, however,
quite involved and its physical content
is often overwhelmed by the technical difficulties found in
applications to realistic cases.

Simple model systems have been proposed to
obtain exact solutions of the Schr\"odinger equation.
These solutions have been used as benchmarks to
test the validity of the various approximations in many-body approaches
\cite{mos68a,mos68b,lat73,nog76,sau78}.
One of these models has been proposed,
in the middle '60s of the last century, by Lipkin,
Meshkov, and Glick (LMG) to describe a many-fermion
system with two energy levels \cite{lip65a,lip65b,lip65c}.

In the spirit of refs.\cite{mos68a,mos68b,lat73,nog76,sau78},
after presenting
a simple derivation of the exact solutions of the LMG
model, we
compare these solutions with the results obtained by using two
approximation methods. The first one, based on the variational principle, is
the Hartree-Fock (HF) theory\cite{har28,foc30,sla50}.
The second one is the Random Phase Approximation (RPA) which was originally
formulated by Bhom and Pines to describe plasma fluctuations\cite{boh52,row68,par68,hag00}.

In section \ref{sec:lipkin}, we discuss the LMG model  and  give, when possible, the analytic
expressions for the energy levels of the system. In section \ref{sec:HF},  we show how the HF approach
can be used to describe the ground state of the system. Then, in section \ref{sec:RPA},
we briefly introduce the RPA theory  and calculate the ground-state beyond the HF
approximation. Our conclusions are drawn in the final section.

%Finally,  we believe that the pedagogical
%presentation of the exact solution of  LMG model (section II), which to the best of our knowledge
%it is a model which, up to now, has been only used in research paper, and the possibility to test the
%validity  of the HF approach (section III) and RPA theory (section IV) by comparing the approximated %solution
%with the exact one  can be useful in advanced courses of Quantum Mechanics, Solid State and Nuclear Physics.

\section{The Lipkin Model}
\label{sec:lipkin}
In its original version~\cite{lip65a}, the LMG model consists of
$N$ fermions occupying two energy levels, each of them has
an $N$-fold degeneracy. We indicate with $\epsilon$ the energy
difference between these two levels. Each level is characterised
by a quantum number  $\sigma$ which assume the value $+1$ in the
upper level and $-1$ in the lower one, and by a set $p$ of
quantum numbers specifying the particular degenerate states
within the same level. Only
two-body interactions which scatter pairs of particles between the two levels
without changing the value of $p$ are considered.
The hamiltonian of this model system is given by
%
%\beq
%H = \epsilon\, K_{0}  - \frac{V}{2}\,\left(K_{+}^{2} + K_{-}^{2}\right) - \frac{W}{2}\, \left\{\,K_{+}\,,\, K_{-}\,\right\},
%\label{eq:hdef}
%\eeq
%
%
\beq
H = \epsilon\, K_{0}  - \frac{V}{2}\,\left(K_{+}^{2} + K_{-}^{2}\right) -
\frac{W}{2}\, \left(K_{+} K_{-}\, + K_{-} K_{+}\, \right),
\label{eq:hdef}
\eeq
%
%\gre{ho scritto esplicitamente $\{,\}$ perch\'e l'anticommutatore non \`e stato ancora definito}
with
\beq
K_{0}= \frac{1}{2}\,\sum_{p=1}^{N}\left( a^{\dag}_{p,+}a_{p,+} -  a^{\dag}_{p,-}a_{p,-} \right),\quad
K_{+}= \sum_{p=1}^{N}a^{\dag}_{p,+}a_{p,-}\quad {\rm and} \quad K_{-}=K_{+}^{\dag}.
\label{eq:def}
\eeq
where $a^\dag_{p,\pm}$ and $a_{p,\pm}$ are the usual fermion creation and annihilation operators
satisfying the anti-commutation relations
\beq
\{\,a_{p,\alpha}\,,\,a^{\dag}_{r,\beta}\,\} = \delta_{pr}\,\delta_{\alpha\beta},\quad
\quad
\{\,a_{p,\alpha}\,,\,a_{r,\beta}\,\} = \{\,a^{\dag}_{p,\alpha}\,,\,a^{\dag}_{r,\beta}\,\} = 0.
\label{eq:acomm}
\eeq
The term proportional to $V$ scatters pair of particles from one level to the other one. The
$K^2_+$ operator removes two particles from the lower level and put them on the upper level.
The $K^2_-$ operator acts in the opposite manner. The term proportional to $W$ promotes a particle
in the upper level, creating a hole in the lower level, and, at the same time, removes a particle
from the upper level and put it in the lower one.

In the unperturbed situation, i.e. when the interactions are switched off ($V=W=0$)
the energies of the various states are given by the number of particles
lying in the upper and lower states, each of them with different values of the quantum
number $p$. The lowest energy, that of the ground state, is obtained when
all the particles lie on the lower level. Each particle--hole excitation produces a new state
whose energy is larger than that of the previous excited state by the
quantity $\epsilon$.

We use the properties of the $K_{0,\pm}$ operators to solve
the Schr\"odinger equation for this many-fermion system.
By using the anticommutation relations (\ref{eq:acomm}), we obtain
\begin{eqnarray}
\left[\,K_{+}\,,\,K_{-}\,\right]
& = &  \sum_{p,r=1}^{N}\,(\,a^{\dag}_{p,+}a_{p,-} a^{\dag}_{r,-}a_{r,+} -  a^{\dag}_{r,-}a_{r,+}a^{\dag}_{p,+}a_{p,-}\,)
\nonumber \\
& = &  \sum_{p,r=1}^{N}\,(\,a^{\dag}_{p,+} a_{r,+} \delta_{pr} -   a^{\dag}_{r,-} a_{p,-}\delta_{rp}   \,)
  =  2 \, K_{0}
\label{eq:kpkm}
\end{eqnarray}
and
\begin{eqnarray}
\left[\,K_{0}\,,\,K_{+}\,\right] & = &\frac{1}{2}\, \sum_{p,r=1}^{N}\, \left[ (\,a^{\dag}_{p,+}a_{p,+} -  a^{\dag}_{p,-}a_{p,-}\,)\,  a^{\dag}_{r,+}a_{r,-} - a^{\dag}_{r,+}a_{r,-}\, (\,a^{\dag}_{p,+}a_{p,+} -  a^{\dag}_{p,-}a_{p,-}\,)\,\right]      \nonumber \\
 & = & \frac{1}{2}\, \sum_{p,r=1}^{N}\, \left(\, a^{\dag}_{p,+}a_{p,+} a^{\dag}_{r,+}a_{r,-} -  a^{\dag}_{r,+}a_{r,-} a^{\dag}_{p,+}a_{p,+} +
 a^{\dag}_{r,+}a_{r,-}a^{\dag}_{p,-}a_{p,-} - a^{\dag}_{p,-}a_{p,-} a^{\dag}_{r,+}a_{r,-}\right)\nonumber \\
  & = & \frac{1}{2}\, \sum_{p,r=1}^{N}\, \left(\, a^{\dag}_{p,+} a_{r,-}\delta_{pr}  +
 a^{\dag}_{r,+}a_{p,-}\delta_{rp} \right)
=   K_{+}.
\label{eq:kzkp}
\end{eqnarray}
From the last commutation relation, we obtain
\begin{equation}
\left[\,K_{0}\,,\,K_{-}\,\right] = -\,\left[\,K_{0}\,,\,K_{+}\,\right]^{\dag} = -\,K_{-}.
\label{eq:kzkm}
\end{equation}
The commutation relations (\ref{eq:kpkm}), (\ref{eq:kzkp}), and (\ref{eq:kzkm}) are those of the
components of an angular momentum operator in quantum mechanics, therefore all the features
related to the angular momentum algebra can be applied to the operators $K_{0,\pm}$.
In particular, we observe that the operators  $K_0$ and
\beq
K^{2} =\frac{1}{2}\, \left\{\,K_{+}\,,\, K_{-}\,\right\} + K_{0}^{2}
\label{eq:ksquare}
\eeq
satisfy $[\,K_0\,,\,K^2\,]=0$ and commute with the hamiltonian (\ref{eq:hdef}).
For these reasons, it is convenient to diagonalise the hamiltonian (\ref{eq:hdef})
in the basis of the eigenstates $ |k,m\rangle$ of $K^2$ and $K_0$, whose properties
are
\beqn
\nonumber
K^2 \,|k,m\rangle & = & k(k+1) \, |k,m\rangle \\
\nonumber
K_0\,|k,m\rangle & = & m\,|k,m\rangle \\
K_{\pm} |k,m\rangle& = & \sqrt{k(k+1)-m(m\pm 1)}\,\,|k,m\pm1\rangle.
\label{eq:keigen}
\eeqn
We observe that the creation and annihilation structure of the $K_0$ operator is such that the $K_0$ eigenvalues correspond
to count the difference between the particles lying on the upper level and those lying on the lower level and then multiply
this difference by a factor 1/2. For example,  when all the $N$ particles are in the upper level
the eigenvalue of $K_0$ is $N/2$, while when they are all in the lower level, the eigenvalue is
$-N/2$. These are the two extreme values that the eigenvalues of $K_0$ can assume. Between these
two eigenvalues there is a discrete sequence of eigenvalues each of them differing by a unit, and this
produce a total number of  $N+1$ states. This is the number of states allowed by the symmetry of the
problem, and therefore the dimension of the matrix to be diagonalized to solve
the eigenvalue problem.

For a given number $N$ of particles, we solve the Schr\"odinger equation
\beq
H |\Psi, k  \rangle = E  |\Psi,  k \rangle
\label{eq:sch}
\eeq
by expressing  the eigenstates $|\Psi,  k \rangle $  as linear combination of
the $| k m \rangle $ states\cite{mes61},
\beq
|\Psi, k \rangle = \sum_{m=-\frac{N}{2}}^{\frac{N}{2}} C_m | k, m \rangle
\label{eq:comb}
\eeq
The problem of finding the $C_m$ coefficients can be expressed in matrix form
by bracketing Eq. (\ref{eq:sch}) with $\langle k, m' | $
\beq
 \sum_{m=-\frac{N}{2}}^{\frac{N}{2}}   C_m  \langle k, m' | H  | k, m \rangle =
   \sum_{m=-\frac{N}{2}}^{\frac{N}{2}} C_m   E_{k,m} \langle k, m' | k, m \rangle = E_{k,m'}  C_{m'}
\eeq
where in the last step we used the orthonormality of the $ | k, m \rangle$ states.
In terms of the quantum numbers $m$ and $k$, the matrix elements
different from zero are
\begin{eqnarray}
\langle k, m | H | k, m \rangle &=&   m\, \epsilon - W [\,k(k+1) - m^{2}\,], \nonumber \\
 \langle k, m | H | k, m +2 \rangle & =&   -\,\frac{V}{2}\,\sqrt{[k(k+1) - m (m-1)][k(k+1)-(m-1)(m-2)]}\,,\\
 \langle k, m + 2 | H | k, m \rangle &=&  \langle k, m | H | k, m +2 \rangle .\nonumber
\end{eqnarray}
For example, the explicit expressions of the hamiltonian
matrix  and its eigenvalues for the systems composed by two and three particles are
\begin{equation}
H_2=\left(\begin{array}{ccc} \epsilon-W & 0 & -V  \\ 0 & -\,2\,W & 0 \\ -\,V & 0 & -\,(\epsilon +W)  \end{array} \right),\quad\quad\quad E_{2}=\left\{\begin{array}{l} +\,\epsilon\, \sqrt{1+  (V/\epsilon)^{2}} - W\\  -\,2\,W\\
 -\,\epsilon\, \sqrt{1+  (V/\epsilon)^{2}} - W \end{array}\right.,
 \nonumber
\end{equation}
and
\begin{eqnarray}
H_3&=&\left(\begin{array}{cccc}
\frac{3}{2}\,(\epsilon-W) & 0 & -\,\sqrt{3}\,\,V & 0  \\
0 &  \frac{1}{2}\,(\epsilon-7\,W) & 0 & -\,\,\sqrt{3}V \\
 -\,\,\sqrt{3}V & 0 &  -\,\frac{1}{2}\,(\epsilon+7\,W)  & 0 \\
 0 & -\,\,\sqrt{3}V & 0 & -\,\frac{3}{2}\,(\epsilon+W)
\end{array} \right),
\nonumber
\end{eqnarray}
\begin{eqnarray}
2\,E_{3}&=&\left\{\begin{array}{l}
 +\,\epsilon\, + 2\,\epsilon \,\sqrt{1+  3\,(V/\epsilon)^{2} + 2 (W/\epsilon)  + (W/\epsilon)^{2}} - 5\,W\\
 -\,\epsilon\, + 2\,\epsilon \,\sqrt{1+  3\,(V/\epsilon)^{2} - 2 (W/\epsilon)  + (W/\epsilon)^{2}} - 5\,W\\
  +\,\epsilon\, - 2\,\epsilon \,\sqrt{1+  3\,(V/\epsilon)^{2} + 2 (W/\epsilon)  + (W/\epsilon)^{2}} - 5\,W\\
   -\,\epsilon\, - 2\,\epsilon \,\sqrt{1+  3\,(V/\epsilon)^{2} - 2 (W/\epsilon)  + (W/\epsilon)^{2}} - 5\,W
\end{array}\right.
\nonumber
\end{eqnarray}
For $W=0$ and $N=4,6,8$ the secular equation is at most quadratic, and exact solutions for the
energy eigenvalues can be obtained analytically~\cite{lip65a},
\begin{eqnarray}
E_4/\epsilon & = & 0,\quad \pm\,\sqrt{1+9\, (V/\epsilon)^2},\quad \pm\,2\, \sqrt{1+3\, (V/\epsilon)^2},\nonumber \\
E_6/\epsilon & = & 0,\quad \pm\,2\, \sqrt{1+15 \,(V/\epsilon)^2},\quad \pm\,\sqrt{5+33 (V/\epsilon)^2\pm 4 \sqrt{1+6 (V/\epsilon)^2+54 (V/\epsilon)^4}},\nonumber \\
E_8/\epsilon & = &   0,\quad \pm\,\sqrt{5+113 (V/\epsilon)^2\pm 4 \sqrt{1+38 (V/\epsilon)^2+550 (V/\epsilon)^4}},
 \nonumber\\
 &  & \quad \quad \pm\,\sqrt{10+118 (V/\epsilon)^2\pm 6 \sqrt{1-2 (V/\epsilon)^2+225 (V/\epsilon)^4}}.
\end{eqnarray}
The above expressions correct those given in the original paper of
Lipkin et al.~\cite{lip65a} where a
factor $4$ in $E_6$ and  $6$ in $E_8$ are missing.

The green full lines of Fig.\,\ref{fig:fig1} and Fig.\,\ref{fig:fig2}
show ground state energy values, $E_{_{\rm GS}}$,
as function of the interaction $V$ for $N=2,3,4$ (Fig.\,\ref{fig:fig1}) and $N=6,8,20$ (Fig.\,\ref{fig:fig2}).
In the left panels we show the solutions obtained when $W=0$, and in the right panels those
for $W=V$. The results up to $N=8$ have been obtained
by using the analytical expressions
shown above, while those for $N=20$
by performing a numerical
diagonalization of the hamiltonian matrix with standard techniques \cite{pre86}.

The solutions we have presented are {\it exact}, meaning that they have been obtained without
making approximations of the problem to be solved. These results are our benchmarks to test the validity
of the simplified solutions of the many-fermion problems presented in the next sections.

\section{Hartree-Fock}
\label{sec:HF}

The HF method\cite{har28,foc30,sla50} is one of the most commonly used approaches to describe
the ground state of many-fermions systems.
The basic HF equations can be obtained in various manners, from
the application of the  Rietz variational principle~\cite{mes61} to the use of the
the first order solution of the Dyson equation in the Green's function expansion method~\cite{fet71}.
In the present paper we consider the HF approach in its variational formulation.

We search for a ground state solution of the LMG fermion system where all the particles occupy the lower
energy level only. This solution is obtained by selecting the low energy level in such a way that the
energy of the system is minimal.

We use the basis of states obtained without interaction $(V=W=0)$, then we switch on the interaction
and we search for the solution which minimises the total energy of the system.
This is equivalent to search for the unitary transformation of the basis of the $| k m \rangle$ states
which generates the {minimal energy solution.

We obtain a unitary transformation of the $| k m \rangle$ basis
by making a unitary trasformation
of the creation and annihilation operators
\begin{equation}
\begin{bmatrix}
	\tilde{a}_{p,+} \\
	\tilde{a}_{p,-}
\end{bmatrix}
=
\begin{pmatrix}
	 \cos \frac{\alpha}{2} & -\sin \frac{\alpha}{2}  \\
	\sin \frac{\alpha}{2}  &  \ \ \, \cos \frac{\alpha}{2}
\end{pmatrix}
\begin{bmatrix}
	a_{p,+} \\
	a_{p,-}
\end{bmatrix}
\,\,,
\end{equation}
where the $a$ operators, and they hermitian conjugate, act on the old basis, while the
$\tilde{a}$ operators on the new basis.
The above transformation defines
the new basis in terms of the states of the old basis.
In the new basis, we can define the operators
\begin{equation}
\kti_{0}= \frac{1}{2}\,\sum_{p=1}^{N}\left( \tilde{a}^{\dag}_{p,+}\tilde{a}_{p,+} -
\tilde{a}^{\dag}_{p,-}\tilde{a}_{p,-} \right)\,\,,\quad
\kti_{+}= \sum_{p=1}^{N}\tilde{a}^{\dag}_{p,+}\tilde{a}_{p,-}\,\,,
\quad {\rm and} \quad \kti_{-}=\kti_{+}^{\dag}
\,\,,
\label{eq:newk}
\end{equation}
which are related to the $K_{0,\pm}$ operators of the old basis by the relation
\begin{equation}
\
\begin{bmatrix}
	\kti_+ \\
	\kti_0 \\
\kti_-
\end{bmatrix}
=\frac{1}{2}\,
\begin{pmatrix}
	 \cos \alpha +1 & \quad 2\,\sin \alpha \quad & \cos\alpha -1 \\
	-\,\sin \alpha  &  2\, \cos \alpha & -\sin \alpha   \\
\cos\alpha -1 & 2\,\sin\alpha & \cos \alpha +1
\end{pmatrix}
\begin{bmatrix}
	K_+ \\
	K_0 \\
K_-
\end{bmatrix}
\,\,.
\label{eq:ktilde}
\end{equation}
In analogy with the $K_{0,\pm}$ operators, the new $\kti_{0,\pm}$ satisfy
the following commutation relations
\begin{equation}
\left[\,\kti_{+}\,,\,\kti_{-}\,\right]
  =  2 \, \kti_{0}\quad {\rm and} \quad \left[\,\kti_{0}\,,\,\kti_{\pm}\,\right]
  =  \pm\, \kti_{+}
  \,\,.
\label{eq:ktcomm}
\end{equation}
We can obtain the expression of the hamiltonian (\ref{eq:hdef}) in the new basis
by inverting the matrix Eq.\,(\ref{eq:ktilde})
\begin{equation}
\begin{bmatrix}
	K_+ \\
	K_0 \\
K_-
\end{bmatrix}
=\frac{1}{2}\,
\begin{pmatrix}
	 \cos \alpha +1 & \quad-\, 2\,\sin \alpha \quad & \cos\alpha -1 \\
	\sin \alpha  & \quad  2\, \cos \alpha & \sin \alpha   \\
\cos\alpha -1 & -\,2\,\sin\alpha & \cos \alpha +1
\end{pmatrix}
\begin{bmatrix}
	\kti_+ \\
	\kti_0 \\
\kti_-
\end{bmatrix}
\,\,,
\end{equation}
and by inserting the above relations in Eq. (\ref{eq:hdef})
\beqn
\hti & = & \frac{\epsilon}{2} \left[ 2\cos \alpha \ \kti_0 + \sin \alpha \ (\kti_+  +  \kti_-)\right] + \frac{W}{2}\left[ \kti^{2}_+  + \kti^{2}_- - \{\,\kti_+\,,\,\kti_- \} \right] \nonumber \\
& &   -\, \frac{V+W}{4} \left[  \sin^2\alpha \ (4\,\kti^2_0 - \{\,\kti_+\,,\,\kti_- \}) -\, \sin 2\alpha \ (\,\{\,\kti_0\,,\,\kti_+ \} +    \{\,\kti_0\,,\,\kti_- \}\,)\right.\nonumber \\
& & \quad \quad \quad \quad \quad  \left. + \   (1 + \cos^2 \alpha)\,(\kti^{2}_+  + \kti^{2}_-  )              \right]
\,\,.
\label{eq:htilde}
\eeqn
To evaluate the expectation value of the hamiltonian Eq. (\ref{eq:hdef}) with respect to the new
state we make use of the following relations
\begin{eqnarray}
\kti_0\,|\tilde{k},\widetilde{m}\rangle & = &
\widetilde{m}\,|\tilde{k},\widetilde{m}\rangle,
\label{eq:kzeigen}
\\
\kti_{\pm} |\tilde{k},\widetilde{m}\rangle& = & \sqrt{\widetilde{k}(\tilde{k}+1)-\widetilde{m}(\widetilde{m}\pm 1)}\,\,|\tilde{k},\widetilde{m}\pm 1\rangle.
\label{eq:kpmeigen}
\end{eqnarray}
Therefore, we obtain
\begin{equation}
\langle\tilde{k},\widetilde{m}|\, \{\,\kti_+\,,\,\kti_- \}\, |\tilde{k},\widetilde{m}\rangle = 2\,[\,\tilde{k}(\tilde{k}+1)-\widetilde{m}^{2}\,]
\,\,.
\end{equation}
In the ground state, we have $\tilde{k}=N/2$ and
$\tilde{m}=-N/2$, consequently
\beq
 \brahf\, \{\,\kti+\,,\,\kti_- \}\,|{\rm HF}\rangle = N
\,\,.
\label{eq:khf}
\eeq
The expectation value of the energy in the new basis can be expressed as,
\beq
E_{\alpha} = \brahf \, \hti\, \kethf
 =   -\,N\,\frac{\epsilon}{2}\left[\cos \alpha \,+ \frac{W}{\epsilon}  + \frac{(N-1)(V+W)}{2\,\epsilon}\,
\sin^2\alpha      \right]
\,\,.
\label{eq:rot}
\eeq
The HF solution is obtained for the values of $\alpha$ which minimize the energy (\ref{eq:rot}),
i.e.
\begin{equation}
\cos \alpha_{_{\rm HF}} = \left\{ \begin{array}{ll} 1 & \quad {\rm for} \quad (N-1)(V+W)<\epsilon\quad {\rm (region\,\,\, I)},  \\  \displaystyle{ \frac{\epsilon}{(N-1)(V+W)}}  &
\quad {\rm for} \quad (N-1)(V+W)>\epsilon\quad {\rm (region\,\,\, II)}.
\end{array}\right.
\end{equation}
By using these values, we obtain for
 the HF energy the expressions
\begin{equation}
E_{_{\rm HF}}= -\,\frac{N}{2}\, \left\{ \begin{array}{ll} \epsilon  + W  & \quad {\rm \,(region\,\,\, I),} \\  \displaystyle{ \frac{\epsilon^2 +(N-1)^2(V+W)^2  }{2\,(N-1)(V+W)}\,+\,W}   &
\quad {\rm (region\,\,\, II).}
\end{array}\right.
\end{equation}
The values of the ground state HF energies as a function of the interaction $V$
are shown by the dash-dotted lines of Fig.\,\ref{fig:fig1} and  Fig.\,\ref{fig:fig2} by dotted lines.
For $W=0$ we observe a remarkable difference with the exact solutions, especially in the region I,
where the HF energies are constant. In the transition point
between the two regions, at   $\epsilon=(N-1)V$, the value of the energy is
 \[-\,\frac{N}{2}\,\,\epsilon
 \,\,.
 \]
For $W=V$ case, we observe a reasonable agreement of the HF solutions with the exact ones, even in the
region I. In this case, at the transition point between the two regions, which is located at
 $\epsilon= 2 (N-1)V$, the value of the HF energy is
 \[ -\,\frac{N\,(2\,N-1)}{4\,(N-1)}\,\,\epsilon
 \,\,.
  \]
In the figures, the thick red points indicates these values.

\section{Random Phase Approximation}
\label{sec:RPA}

The second effective theory we consider is the RPA, which was originally formulated to
describe the excitations of an electron gas induced by
plasma fluctuations \cite{boh52}, and
in the following has been widely applied to describe
harmonic vibrations of many-fermion systems from atoms to nuclei \cite{row68,par68,hag00}.
The main goal of the RPA theory is the description of the excited states of the system, but
the theory is based on an ansatz about the ground state
which is more elaborated than that used in HF.

We present the basic steps required by the RPA theory to obtain an expression of the ground state energy.
The starting point is the definition of the operator
\begin{equation}
Q^{\dag} = \frac{X\,\kti_{+} -Y\,\kti_{-} }{ \sqrt{N}}
\label{eq:qdag}
\end{equation}
which applied to the RPA ground state $\ketrpa$ describe the excited state
$\keto$
\beq
Q^{\dag} \, \ketrpa =  \keto
\label{eq:rpa1}
\eeq
The $Q^\dag$ operator generates a linear combination of
one-particle one-hole, and one-hole one-particle, excitations.
In the LMG model this ansatz produces only one excited state which we have
identified with $\keto$. The RPA ground state is defined by the equation
\beq
Q\,|0\rangle = 0,
\label{eq:rpa0}
\eeq
evidently $\ketrpa$ cannot be $\kethf$ as the expression
(\ref{eq:khf}) indicates.
The above definitions and the Schr\"odinger equation imply
\beqn
\hti\,Q^{\dag}\, |0\rangle &=& \erpa_1\, |1\rangle,\\
Q^{\dag}\, \hti\, |0\rangle  & = & \erpa_0\,Q^{\dag}\,  |0\rangle,
\eeqn
therefore
\beq
\left[\,\hti\,,\,Q^{\dag}\,\right]\, |0\rangle = (\erpa_1-\erpa_0)\, Q^{\dag}\,|0\rangle =\omega\, Q^{\dag}\,|0\rangle
\eeq
which defines the excitation energy $\omega$. By using the previous equation we write
\begin{eqnarray}
\nonumber
 \brarpa \, \left[\,
   \kti_{-}\,,\,\left[\,\hti\,,\,Q^{\dag}\,\right]\,\right]
 \, \ketrpa  & = & \omega\, \brarpa \,  \left[\,
   \kti_{-}\,,\,Q^{\dag}\,\right] \, \ketrpa,
 \\
 \brarpa \left[\,
   \kti_{+}\,,\,\left[\,\hti\,,\,Q^{\dag}\,\right]\,\right]
 \, \ketrpa  & = & \omega\,  \brarpa \left[\,
   \kti_{+}\,,\,Q^{\dag}\,\right] \, \ketrpa.
\label{eq:rpa2}
\end{eqnarray}
The evaluation of the matrix elements of the above equations is rather
difficult since the RPA ground-state defined by Eq. (\ref{eq:rpa0}) is
not defined in terms of particle-hole excitations in our basis. For
this reason, the so-called Quasi Boson Approximation (QBA) is used
\cite{rin80}. This approximation consits in substituting all the RPA
matrix elements containing commutators of particle - hole pair
operators with the values obtained by calculating them between HF
ground state, i.e.
\beq
\brarpa \,\left[a^+_h a_p,a^+_{p'} a_{h'}\right]\,\ketrpa
\simeq
\brahf \,\left[a^+_h a_p,a^+_{p'} a_{h'}\right] \, \kethf
=\delta_{pp'} \, \delta_{hh'}
\label{eq:QBA}
\eeq
The name Quasi Boson Approximation is given since the operator pair
$a^+_{p'} a_{h'}$
behaves as a single boson operator. In the specific case under investigation, we have that
\beq
\brarpa \,\left[\kti_-,\kti_+ \right]\,\ketrpa
\simeq
\brahf \,\left[\kti_-,\kti_+ \right]\,\kethf
=-\,2 \,\brahf \, \kti_0 \,\kethf
= N,
\label{eq:QBAk}
\eeq
where we used the commutation relations (\ref{eq:ktcomm}) the relation
(\ref{eq:kzeigen}) and the fact that in the HF ground state
$\kethf$ we have $\tilde{m}= - N/2$.
Evidently, the result of Eq. (\ref{eq:QBAk}) indicates that the
$\kti_\pm$ operators between HF states commute as boson operators.

The RPA master equations Eq. (\ref{eq:rpa2})
are commonly written by introducing the quantities
\beqn
A  =  \frac{ \brarpa \, [\,\kti_{-}\,,\, [\,\hti\,,\,\kti_{+}\,]\,] \,
  \ketrpa }{ N}
&\simeq&
\frac{ \brahf \, [\,\kti_{-}\,,\, [\,\hti\,,\,\kti_{+}\,]\,] \,
  \kethf }{ N}
\label{eq:A}
\,\,,
\\
 B =  -\,\frac{ \brarpa \, [\,\kti_{-}\,,\, [\,\hti\,,\,\kti_{-}\,]\,] \, \ketrpa }{N}
&\simeq&
  -\,\frac{ \brahf \, [\,\kti_{-}\,,\, [\,\hti\,,\,\kti_{-}\,]\,] \, \kethf }{N}
\,\,,
\label{eq:andb}
\eeqn
where we have already indicated the use of the QBA which allows us to rewrite Eqs.(\ref{eq:rpa2}) in matrix form
\begin{equation}
\begin{pmatrix}
	 A - \omega & B \\
	B^{*}  & A^* + \omega
\end{pmatrix}
\begin{bmatrix}
	X\\
Y
\end{bmatrix}=0.
\end{equation}
Observing that $A=A^*$ and solving the eigenvalue problem, we find
\beq
\omega = \sqrt{A^{2}-|B|^{2}}\quad\quad\quad{\rm and}\quad\quad\quad  X=\frac{B}{\omega-A}\,Y
\label{eq:xanda}
\eeq
The problem is completely defined after fixing the overall constant
relating $X$ and $Y$. In QBA we have
\beq
\brarpa [\,Q\,,\,Q^{\dag}\,]\ \ketrpa \simeq
\brahf \,[\,Q\,,\,Q^{\dag}\,]\, \kethf =1
\eeq
which implies the following normalization condition
\beq
|X|^{2}-|Y|^{2}=1.
\label{eq:xycomp}
\eeq
By using the properties (\ref{eq:rpa1}) and (\ref{eq:rpa0})
we have that,
\beqn
\brao \, Q^{\dag} \, \ketrpa =1 & \quad\Rightarrow \quad  &
X\, \brao \, \kti_+ \, \ketrpa  -  Y\, \brao \, \kti_- \,\ketrpa =\sqrt{N}
\,\,,
\\
\brao \, Q\, \ketrpa =0 & \quad\Rightarrow \quad  &
X^*\, \brao \, \kti_- \,|0\rangle   =  Y^*\, \brao \, \kti_+ \, \ketrpa
\,\,,
\eeqn
and
\begin{equation}
\brao\, \kti_+ \, \ketrpa =
\sqrt{N}\,X^*\quad\quad {\rm and}\quad\quad
\brao\, \kti_- \, \ketrpa =\sqrt{N}\,Y^*
\,\,.
\label{eq:xandy}
\end{equation}
Many terms of $\hti$ do not contribute to the
expectation value with respect to the RPA ground state. For this reason,
instead of carring out the calculation
with the full hamiltonian $\hti$, we use an effective hamiltonian
whose terms generate contributions
 different from zero and satisfy
Eq.~(\ref{eq:rpa1}). The general expression of this hamiltonian,
up to quadratic terms in $\kti$, is given by \cite{rin80}
\beq
H_{\rm RPA} =  E_{\rm HF}
+ \frac {1}{N} \left[ A \kti_+ \kti_- +
             \frac{1}{2} \left( B \kti^2_+ + B^* \kti^2_- \right)
               \right]
\label{eq:hrpa}
\eeq
We can use the effective RPA hamiltonian (\ref{eq:hrpa}) to calculate the energy of the RPA ground state
\begin{eqnarray}
E_{_{\rm RPA}} & = &  \langle 0 |\, H_{_{\rm RPA}}\,|0\rangle\nonumber\\
  & = & E_{_{\rm HF}}+  \frac{A\,\langle 0 |\,\kti_+\kti_-   \,|0\rangle}{N}+ \frac{\langle 0 |\, (\,B\,\kti_+^{2} + B^*\kti_-^{2}\,) \,|0\rangle}{2\,N}\nonumber \\
   & = & E_{_{\rm HF}}+  \frac{A\,\langle 0 |\,\kti_+ \,|1\rangle\langle 1| \,\kti_-   \,|0\rangle}{N}+ \frac{B\,
  \,\langle 0 |\,\kti_+ \,|1\rangle\langle 1| \,\kti_+   \,|0\rangle +B^*
  \langle 0 |\,\kti_- \,|1\rangle\langle 1| \,\kti_-   \,|0\rangle}{2\,N}\nonumber \\
 & = & E_{_{\rm HF}}+  A\,|Y|^{2} + \frac{B\,Y\,X^* + B^*Y^*X}{2}\nonumber \\
  & = & E_{_{\rm HF}}+ \left(A + \frac{|B|^{2}}{\omega-A}\right) |Y|^2\nonumber \\
  & =&  E_{_{\rm HF}} - \omega\,  |Y|^2
  \,\,.
\end{eqnarray}
Considering the relation (\ref{eq:xanda}) and the normalisation of the $X$ and $Y$
(\ref{eq:xycomp}), we have
\beq
 |Y|^2 =\frac{A-\omega}{2\,\omega}
\eeq
and we can write the energy of the RPA ground state as
\begin{equation}
\erpa_0 =  E_{_{\rm HF}} + \frac{\omega-A}{2}
\label{eq:rpa}
\end{equation}
The evaluation of the the $A$ and $B$ coefficients is carried out in QBA.
By using the commutation relations (\ref{eq:ktcomm}) between the
$\kti_{0,\pm}$ operators we obtain
\beq
\langle  {\rm HF}  | \,[\,\kti_{-}\,,\, [\,\kti_{0}\,,\,\kti_{+}\,]\,]\, |  {\rm HF} \rangle =\langle {\rm HF}  |\,  [\,\kti_{-}\,,\,\kti_{+}\,]\,| {\rm HF} \rangle =N
\eeq
Observing that
\[
[\,\kti_{-}\,,\, [\,\kti^2_{0}\,,\,\kti_{+}\,]=\mbox{$\frac{1}{2}$}\,
[\,\kti_{-}\,,\, [\,\kti^2_{+}\,,\,\kti_{-}\,]\,]=-\,\mbox{$\frac{1}{2}$}\,
[\,\kti_{-}\,,\, [\,\{\,\kti_+\,,\,\kti_- \}\,,\,\kti_{+}\,]\,]
\]
and
\[ \langle\tilde{k},\widetilde{m}|\,[\,\kti_{-}\,,\, [\,\kti^2_{0}\,,\,\kti_{+}\,]\,]\,
|\tilde{k},\widetilde{m}\rangle =2\,[\,\tilde{k}(\tilde{k}+1) -3\,\widetilde{m}^{2}\,]  \]
we find
\begin{equation}
\begin{array}{ccr}
\langle  {\rm HF}|\,[\,\kti_{-}\,,\, [\,\kti^2_{0}\,,\,\kti_{+}\,]\,]\,|  {\rm HF}\rangle   & = & -\, N(N-1)\\
\langle  {\rm HF}|\,[\,\kti_{-}\,,\, [\,\{\,\kti_+\,,\,\kti_- \}\,,\,\kti_{+}\,]\,]\,| {\rm HF}\rangle   & = & 2\,N(N-1)\\
\langle  {\rm HF}|\,[\,\kti_{-}\,,\, [\,\kti^2_{+}\,,\,\kti_{-}\,]\,]\,|  {\rm HF}\rangle   & = & -\,2\, N(N-1)\\
\end{array}
\end{equation}
Putting together the previous results with the expression of the hamiltonian
(\ref{eq:htilde}), we obtain
\begin{eqnarray}
A & = & \epsilon\,\cos\alpha + \frac{3}{2}\,(N-1)\,(V+W)\,\sin^2\alpha - (N-1)\,W,\nonumber \\ \\
B & = & -\, (N-1)\,(V+W)\,\frac{1+\cos^2\alpha}{2} + (N-1)\,W. \nonumber
\end{eqnarray}
For $\alpha=\alpha_{_{\rm HF}}$, we then find
\begin{equation}
A =  \left\{ \begin{array}{ll} \epsilon  - (N-1)\,W & \quad {\rm (region\,\,\,I,),} \\   \displaystyle{ \frac{3\,(N-1)^2(V+W)^2  - \epsilon^2   }{2\,(N-1)(V+W)}\, - (N-1)\,W }     &
\quad {\rm (region\,\,\,II),}
\end{array}\right.
\end{equation}
and
\begin{equation}
 B=  \left\{ \begin{array}{ll}  - \,(N-1)\,V & \quad {\rm \,(region\,\,\,I),} \\   \displaystyle{ -\,\frac{\epsilon^2  +  (N-1)^2(V+W)^2}{2\,(N-1)(V+W)}\, + (N-1)\,W }     &
\quad {\rm (region\,\,\,II).}
\end{array}\right.
\end{equation}
The behaviour of the RPA ground state energies, as a function of the strength $V$
of the interaction is shown in Figs.\,\ref{fig:fig1} and \ref{fig:fig2} by the blue dashed lines.
For $W=0$, it is evident the improvement with respect to the HF results, especially in
the region I. The value of the energy at the transition point between the two regions is
\[ -\,\frac{N+1}{2}\,\epsilon
\,\,.\]
For $W=V$ the agreement between RPA and exact results in the region I is excellent.
In this case the value of the energy in the transition point is
\[  - \,\frac{4\,N^{2}-1}{4\,(N-1)}\,\,\epsilon
\,\,.   \]
The behaviour of the solutions for $W=V$ in the region II is remarkable. In this
region we find for the RPA solution
\begin{equation}
B=  -\,\frac{\epsilon^2}{4\,(N-1)V}
\,\,.
\end{equation}
For $4\,(N-1)V \gg \epsilon$ we have that $B \to 0$
and, consequently, due to the fact that $Y \to 0$, the value RPA energy $\erpa_0$
tends to that of the HF energy.

\section{Conclusions}
\label{sec:con}
The LMG many-fermion model, composed by two energy levels,
is an ideal system where the
many-body Schr\"odinger equation for the interacting particles can be solved without
approximations. The comparison of these exact solutions with those obtained by using
effective theories can give a measure of the validity of the latter ones.

We considered a hamiltonian containing two interacting terms. A first one, whose strength has
been called $V$, scatters pairs of particles from one level to the other one, and a second term,
whose strength is $W$, removes a particle from one level and put it on the
other one.

In this article, we tested the validity of the HF and RPA theories in the description of the ground state
of the system. In both cases, the solutions are characterised by two regions which depend on the strength of the interaction between the particles. The transition between the two regions is discontinuous. The discontinuity at the meeting point (which seems to suggest some sort of phase transition) is clearly an artefact of the effective theories, since the exact results do not present any discontinuity region. So, approximation methods could suggests anomalous behaviors which do not really happen in the real system. For strong interactions and large number of interacting particles the HF and RPA solutions approach the exact  behavior of the system.

In the region I, characterised by relatively small values of the interaction,
the HF energies are independent of $V$. This implies that for $W=0$, these energies
are constant and show a remarkable discrepancy with respect to the
exact result which become smaller when $V$ increases.
The HF description of the exact results improves in the region II, and also
when $W=V$.

The RPA solutions give a reasonable description of the exact results also in the region I for $W=0$,
showing a large improvement with respect to the HF results.
In all the cases we have considered, the RPA and HF
solutions converge for large values of the interaction strength $V$.
The solutions of the effective theories become closer to the exact ones when the number of
the particles composing the system increases.

%\begin{acknowledgments}
%\end{acknowledgments}

%\newpage

\newpage

\begin{figure}
\vspace*{-1.5cm}
\hspace*{-1.2cm}
\includegraphics[scale=0.75]{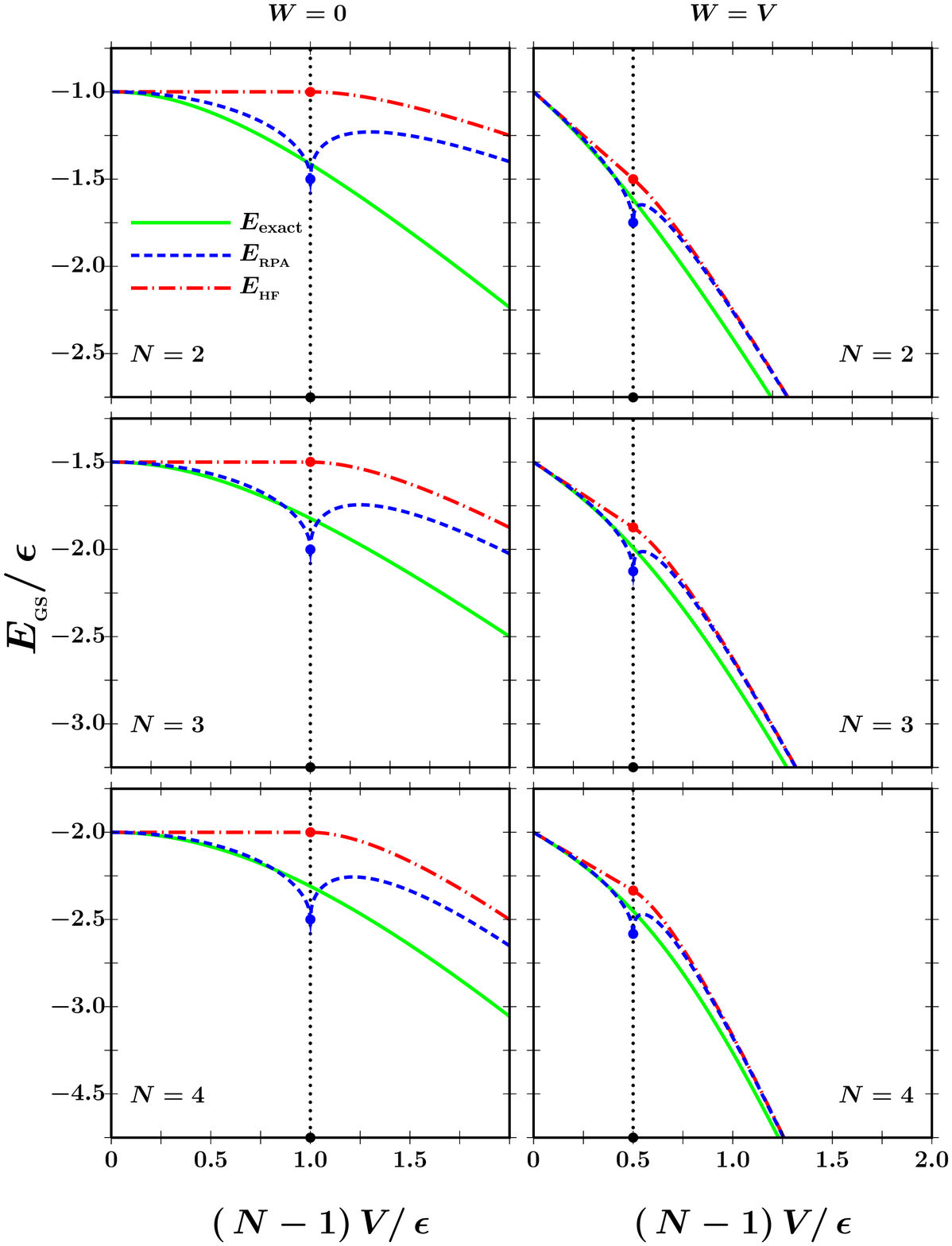}
%\vspace*{-2cm}
\caption{Ground state energies as a function of the interaction strength $V$ for systems
composed by $N=2,3$ and 4 particles. The full (green) lines indicate the solutions
obtained without approximations. The dash-dotted (red) lines the results obtained with the HF
model, and the dashed (blue) lines those obtained with the RPA approach. The left panels
show the results for $W=0$, and the right panels those obtained by setting $W=V$.
The blue and red thick dots emphasize the values of the HF and RPA energies in the discontinuity
line.
}
\label{fig:fig1}
\end{figure}

\begin{figure}
\vspace*{-1.5cm}
\hspace*{-1.2cm}
\includegraphics[scale=0.75]{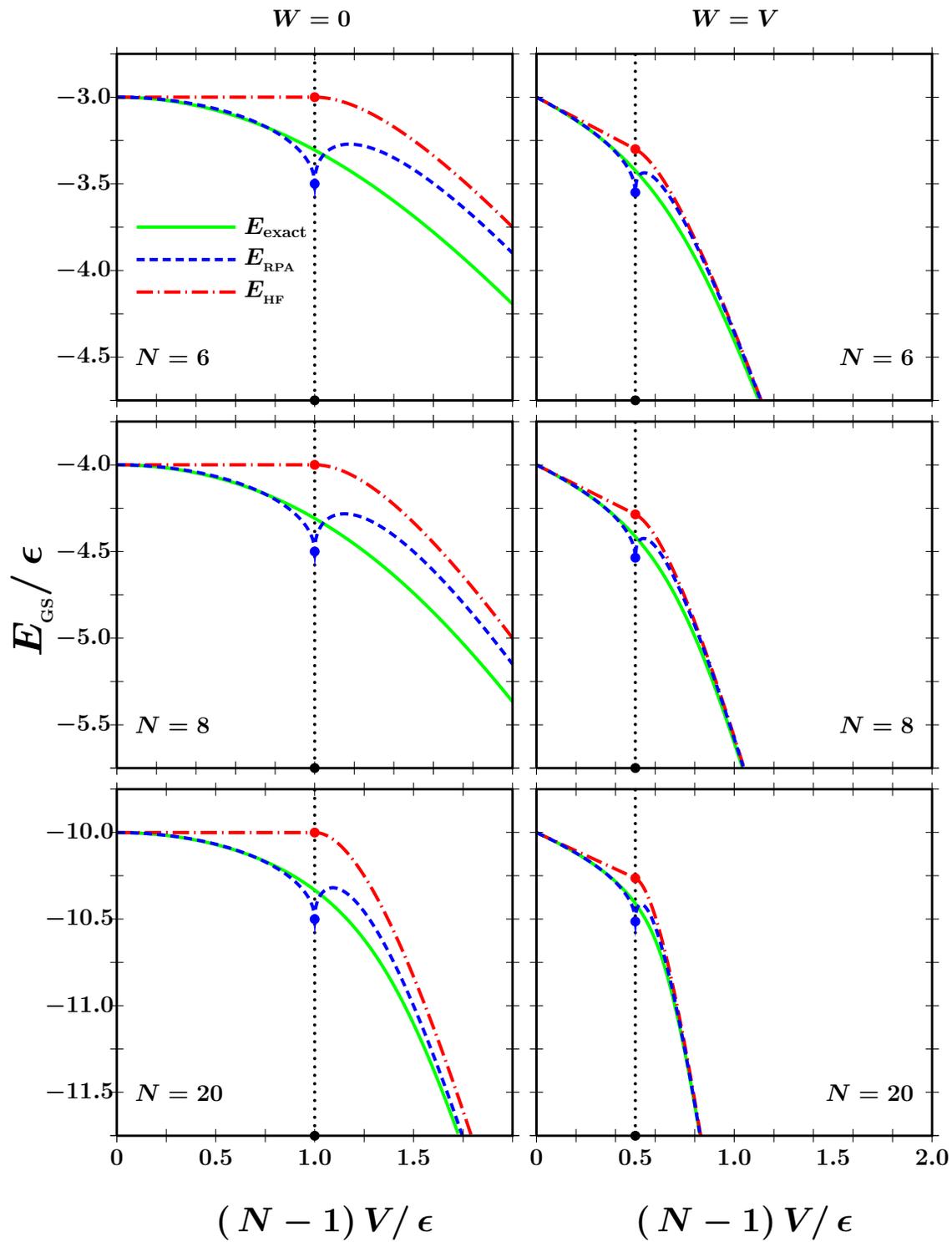}
%\vspace*{-1.5cm}
\caption{The same as in Fig. \ref{fig:fig1} for $N=6,8$ and 20.}
\label{fig:fig2}
\end{figure}

\end{document}